\documentclass[12pt]{article}

\usepackage[colorlinks,linkcolor=blue,citecolor=blue,urlcolor=blue,bookmarks,bookmarksnumbered]{hyperref}
\usepackage{amsmath,amssymb,amsfonts}
\usepackage[paper=letterpaper,margin=1.0in]{geometry}
\usepackage{graphicx,xcolor}
\usepackage{braket}

\usepackage{cite}

\newcommand{\be}{\begin{equation}}
\newcommand{\ee}{\end{equation}}
\newcommand{\bea}{\begin{eqnarray}}
\newcommand{\eea}{\end{eqnarray}}
\newcommand{\ba}{\begin{array}}
\newcommand{\ea}{\end{array}}
\newcommand{\ben}{\begin{enumerate}}
\newcommand{\een}{\end{enumerate}}
\newcommand{\bi}{\begin{itemize}}
\newcommand{\ei}{\end{itemize}}
\newcommand{\bc}{\begin{center}}
\newcommand{\ec}{\end{center}}
\newcommand{\bfig}{\begin{figure}}
\newcommand{\efig}{\end{figure}}
\newcommand{\bq}{\begin{quotation}}
\newcommand{\eq}{\end{quotation}}
\newcommand{\bt}{\begin{table}}
\newcommand{\et}{\end{table}}
\newcommand{\btab}{\begin{tabular}}
\newcommand{\etab}{\end{tabular}}
\newcommand{\bs}{\begin{slide}}
\newcommand{\es}{\end{slide}}

\newcommand{\IR}{\mathbb{R}}

\let\ba=\overline

\def\IR{\relax\leavevmode{\rm I\kern-.18em R}}
\def\ZZ{\relax\leavevmode
       \ifmmode\mathchoice
       {\hbox{\sf Z\kern-.4em Z}}
       {\hbox{\sf Z\kern-.4em Z}}
       {\lower.9pt\hbox{\scriptsize\sf Z\kern-.36em Z}}
       {\lower1.2pt\hbox{\tiny\sf Z\kern-.36em Z}}
       \else{\sf Z\kern-.4em Z}\fi}
\def\RR{\relax\leavevmode
       \ifmmode\mathchoice
       {\hbox{\sf R\kern-.4em R}}
       {\hbox{\sf R\kern-.4em R}}
       {\lower.9pt\hbox{\scriptsize\sf R\kern-.36em R}}
       {\lower1.2pt\hbox{\tiny\sf R\kern-.36em R}}
       \else{\sf R\kern-.4em R}\fi}

\def\resetby#1#2{\@addtoreset{#2}{#1}}
\def\seceq{\@addtoreset{equation}{section}
              \def\theequation{\thesection.\arabic{equation}}}

\def\Label#1{\label{#1}%
                \smash{\hbox to0pt{\raise1ex\hbox{\tiny[#1]}\hss}}}
\def\noLabels{\let\Label=\label}

\begin{document}

{\footnotesize
${}$
}

\bc

\vskip 1.0cm

{\Large \bf 
Gravitizing the Quantum}\\

\vskip 1.0cm

\renewcommand{\thefootnote}{\fnsymbol{footnote}}

\bf Per Berglund${}^{1}$\footnote{Per.Berglund@unh.edu},  
\bf Tristan H{\"u}bsch${}^{2}$\footnote{thubsch@howard.edu}
{\bf  David Mattingly${}^{1}$\footnote{David.Mattingly@unh.edu} and} 
{ \bf Djordje Minic${}^{3}$\footnote{dminic@vt.edu (Corresponding Author)} } \\

\vskip 0.5cm

{\it
${}^1$Department of Physics and Astronomy, University of New Hampshire, Durham, NH 03824, U.S.A. \\
${}^2$Department of Physics and Astronomy, Howard University, Washington, D.C.  20059, U.S.A. \\
${}^3$Department  of Physics, Virginia Tech, Blacksburg, VA 24061, U.S.A. \\
}

\ec

\vskip 1.0cm

\begin{abstract}
We discuss a new approach to the problem of quantum gravity in which
the quantum mechanical structures that are traditionally fixed, such as the Fubini-Study metric in the Hilbert space of states, become dynamical and so implement the idea of {\it gravitizing the quantum}. 
In particular, in this formulation of quantum gravity the quantum geometry is still consistent with the principles of unitarity and  also captures  fundamental aspects of (quantum) gravity, such as topology change.  
Furthermore, we  address specific ways of testing this new approach to quantum gravity by utilizing multi-path interference and optical lattice atomic clocks.

\end{abstract}

\vspace{0.5cm}

\begin{center}

Essay written for the Gravity Research Foundation \\

2022 Awards for Essays on Gravitation.\\

Submission date: March 30, 2022 \\

\end{center}

\vspace{0.5cm}

\renewcommand{\thefootnote}{\arabic{footnote}}

\newpage

How to fully unify the underlying principles of quantum theory with those of general relativity --- the problem of quantum gravity ---
is still an outstanding question in physics.
After almost a century of intense and diverse theoretical research, there are many approaches  that address this fundamental problem. This large class of models,  
coupled with the historic discovery of gravitational waves and the emergence of new experimental techniques from a variety of
research programs, has led to the new field of quantum gravity phenomenology~\cite{Addazi:2021xuf}. 
This burgeoning field %
 encompasses everything from searches for spacetime symmetry violation in high energy astrophysics to proposals that employ quantum information techniques and nanomechanical oscillators to show empirically that gravity is quantized\cite{Bose:2017nin, Marletto:2017kzi}.

In this essay, we  discuss a new approach to the problem of quantum gravity, based on the idea of {\it gravitizing the quantum}~\cite{Berglund:2022qcc}, by extending the dynamical aspects of general covariance, which  postulates that all quantities in our physical theories must be dynamical, to quantum mechanical structures.\footnote{Recall that   
key to the development of classical general relativity is the notion of general covariance, which can be split into two distinct concepts~\cite{Norton}. The first, {\em\/coordinate invariance,} can be applied to almost any physical theory --- one can for example construct a coordinate invariant formulation of Newtonian gravity~\cite{Norton}.  The second, that of {\em\/dynamicism,} is much more profound as it postulates that all quantities in our physical theories must be dynamical. It is this that takes us from the fixed stage of Newtonian gravity and special relativity into the full framework of general relativity.}
 We propose that quantum geometry (by which me mean the actual geometry typically imposed on Hilbert spaces by quantum mechanical axioms and not a quantized spacetime geometry) becomes dynamical in quantum gravity in a way that is consistent with the principles of unitarity and that captures some fundamental aspects of gravity, such as topology change.  We also address specific ways of testing this new approach to quantum gravity by utilizing multi-path interference and optical lattice atomic clocks.\footnote{
Similar ideas have been discussed by Penrose as a way to address the quantum measurement problem~\cite{penrose} and in attempts to understand deeper axiomatic foundations of quantum theory
and quantum gravity~\cite{Hardy:2001jk}.}

To understand how quantum geometry might become dynamical, consider the case of \hbox{$0{+}1$}-dimensional quantum gravity, which is defined by the action for one dimensional universes~\cite{Strominger:1988ys},
\be
S_P= \int d\tau ( e^{-1} \dot{x}^a \dot{x}^b g_{ab} - e\, m^2),
\ee
i.e., scalar particles with mass $m$. Here $\tau$ denotes the worldline proper time,
and the einbein $e(\tau)$ captures the worldline geometry (``gravity'').
We now fix the gauge, $e(\tau) =N$, with $N$ being the lapse.  In this gauge the classical momentum is
 $p_b = \frac{1}{N} \dot{x}^a g_{ab}$,
or $p_a = - i \frac{\partial}{\partial x^a}$ in the quantum context, with the Hamiltonian given by $H= g^{ab} p_a p_b + m^2$.
The Hamiltonian constraint (implied by the underlying diffeomorphism invariance) leads to the Wheeler-DeWitt (``Einstein-Schr\"{o}dinger'') equation
$H\psi(x) =0$, where $\psi(x)$ is the wave-function of the $0{+}1$-dimensional universes. In the absence of any interactions (topology change for one-dimensional universes) this equation entirely captures the
quantum dynamics of this so-called first-quantized formulation. Because of its linearity one has the canonical quantum interference phenomena displaying
the superposition of quantum states.

From the $D$-dimensional spacetime target space point of view, one could view the above Wheeler-DeWitt equation as a classical equation of motion for a free scalar field,
 obtained by varying the following canonical Klein-Gordon-like action
 $S_{\psi} = \int d^{D}\!x\, \sqrt{-g} \frac{1}{2} \psi H \psi$. 
 This is sometimes called the second-quantized formulation. In essence,
\be
S_{\psi} = \int d^{D}\!x\, \sqrt{-g}~ \frac{1}{2} \psi H \psi ~\to~ H \psi(x) =0.
\ee

Now, imagine including topology change (such as a cubic vertex for $\psi$) or, in other words, interactions.
This so-called third-quantized formulation leads to the interacting action $S_{\psi}(g)= \int d^{D}\!x\,\sqrt{-g}\; [\frac{1}{2}\psi H \psi + \frac{g}{3} \psi^3]$,
where $g$ is the coupling constant associated with the cubic vertex.
The variation of this action gives a non-linear extension of the Wheeler-DeWitt equation:
\be
S_{\psi}(g)= \int d^{D}\!x\,\sqrt{-g}~ \Big[\frac{1}{2}\psi H \psi +\frac{g}{3}\psi^3\Big] 
 ~\to~ H \psi(x) + g \psi^2(x)=0.
\ee

The failure of the Hamiltonian constraint to vanish in the above reflects the underlying breakdown of diffeomorphism invariance in this formulation, which may give the reader pause as the underlying Hilbert space of quantum gravity, if not anomalous, should maintain the symmetries of the classical theory.  However, one should not expect diffeomorphism invariance to hold exactly along the worldline in our example as there can exist no 
diffeomorphism between different worldline topologies (or more generally between different spacetimes).   
Thus, quantum gravity effects have made the Schr\"{o}dinger equation of one dimensional universes dynamical; it has been ``gravitized'' in the presence of the above topology change.\footnote{Topology and topology change in quantum gravity has been studied by many authors, e.g.~\cite{Hawking:1979pi,Dowker:2002hm,Horowitz:1990qb}, however the details are not relevant for this essay.}

To summarize: the first quantization is quantum field theory from the viewpoint of $0{+}1$-dimensional quantum gravity.
Second quantization is the Wheeler-DeWitt equation for this worldline formulation.
Third quantization then 
includes interactions, i.e., topology change 
from the worldline point of view.
The essential point is that third quantization requires a non-linear Wheeler-DeWitt equation, and thus a dynamical or ``gravitized''
quantum evolution from the worldline viewpoint.  Note that the target space description is canonical: one has an interacting (cubic) scalar field theory and the usual, canonical formulation
of its quantization of a target space quantum field theory, which is linear and unitary.
The new insight is that the cubic vertex of the cubic scalar field theory (viewed as a non-linear Wheeler-DeWitt equation)
already contains the gravitization of the Schr\"{o}dinger equation, by making it non-linear.

How would a $0{+}1$-dimensional observer test the non-linear Wheeler-DeWitt equation? Without topology change an observer could have constructed (in some gauge) a conserved energy.  The topology change would then manifest itself as a loss of energy for an observer who remains on a single branch as implied by the wave-function dependent shift of the Hamiltonian
\be
H_{\psi}\equiv H + g\psi(x), \quad H_{\psi}\,\psi =0.
\ee
Note that unitarity is
maintained from the target space point of view, while  from the $0{+}1$-dimensional perspective the formal ``unitary'' operator becomes
wave-function dependent. In the limit of small $g$ coupling,
\be
U_{\psi} \equiv \exp\{iH_{\psi} t\} = \exp\{i (H+ g \psi) t\},
\label{Upsi}
\ee
where in a time-reparametrization invariant theory $t$ is defined in a specific gauge, which is usually achieved
by an introduction of an explicit ``clock'' variable in the system under consideration.\footnote{Note that we are not concerned with
the general conceptual ``problem of time'' in this presentation, but only its practical aspect, as discussed in~\cite{Bojowald:2010xp}.}

An even more interesting conceptual (experimental) consequence of the gravitization of the Schr\"{o}dinger equation would be
an intrinsic triple-path interference given the cubic nature of
the discussed topology change.\footnote{Such higher order, non-linear interference patterns in $n$-slit quantum experiments have been discussed 
in a different context by Sorkin~\cite{Sorkin:1994dt}.}  
In order to see how, 
let us denote by
$
P_{n}(A,B,C,\cdots)
$
the probability of a system to go from an initial state $\ket{\alpha}$ to a final state $\ket{\beta}$
when $n$ pathways $A,B,C,\dots$ connecting the two are available, following the presentation in~\cite{Huber:2021xpx}.
Classically, we have
\be
P_{n}(A,B,C,\cdots) \,=\, P_{1}(A) + P_{1}(B) + P_{1}(C) + \cdots\;,
\ee
for any number of paths.
Quantum mechanically, we have for two paths
$
P_{2}(A,B) = |\psi_A + \psi_B|^2 \vphantom{\Big|}$
or more explicitly
\be
{|\psi_A|^2} + 
{|\psi_B|^2} + 
{(\psi_A^*\psi_B^{\phantom{*}} + \psi_B^*\psi_A^{\phantom{*}})}
\equiv P_{1}(A) + P_{1}(B) + I_{2}(A,B)
\ee
%
where the last term
\be
I_{2}(A,B) = P_{2}(A,B)-P_{1}(A)-P_{1}(B) 
\ee
is the ``interference'' of the two paths $A$ and $B$.
The non-vanishing of this double-path interference, 
$I_{2}(A,B)\neq 0$, distinguishes quantum theory from the classical one.
The Born rule dictates that all the superimposed paths only interfere with each other 
in a pairwise manner.
For instance, for three paths we have
\be
P_{3}(A,B,C) = |\psi_A {+} \psi_B {+} \psi_C|^2 \equiv  P_{2}(A,B) {+} P_{2}(B,C) {+} P_{2}(C,A) 
{-} P_{1}(A) {-} P_{1}(B) {-} P_{1}(C),
\label{eq:three-slit}
\ee
where only pairwise interferences between the pairs $(A,B)$, $(B,C)$, and $(C,A)$ appear.

It is clear from the above that in order for there to be a non-linear correction in an interference pattern the Born rule must be relaxed.  However, our example of topology change indicates that this should in fact occur.  The Born rule for $P_{n}(A,B,C,\cdots)$ can be written as
\be
P_{n}(A,B,C,\cdots)=|\braket{\beta |U_\alpha\, \alpha}|^2
\ee
for the state-dependent unitary evolution operator $U_\alpha$, as introduced in~\eqref{Upsi}, that acts on state $\alpha$ and evolves it to the time at which  $\beta$ is measured.  This has an equivalent ``time-reversed'' formulation  
\be
P_{n}(A,B,C,\cdots)=|\braket{\alpha |U^\dagger_\beta\, \beta}|^2
\ee
If the evolution operator $U$ is the usual linear operator independent of the state it acts on then these two expressions are equivalent.  However, in the case of non-linear evolution it matters which state the evolution operators $U_\alpha$ and $U_\beta$ act on.  Hence $U_\alpha$ and $U_\beta$ will generically not match, and the two expressions for $P_{n}(A,B,C,\cdots)$ will differ, leading to an ambiguity at best and a contradiction at worst.  The contradiction can be evaded if the Born rule itself is modified by additional non-linear terms when calculating probabilities (cf.~\cite{Helou and Chen}).  The modified Born rule would then generate deviations from the expected interference patterns in quantum mechanics.

Sorkin has discussed intrinsic higher order interferences in~\cite{Sorkin:1994dt}.  Consider a triple slit experiment
as in equation~\eqref{eq:three-slit}.
Since only pairwise interferences between the pairs $(A,B)$, $(B,C)$, and $(C,A)$ appear, it makes sense to define any deviation from this relation as the intrinsic
triple-path interference~\cite{Sorkin:1994dt} 
\be
I_{3}(A,B,C) \overset{\scriptscriptstyle\text{def}}=
P_{3}(A,B,C)
-P_{2}(A,B)
-P_{2}(B,C)
-P_{2}(C,A)
+P_{1}(A)
+P_{1}(B)
+P_{1}(C).
\ee
(This can be easily generalized for the case of $n$-paths.)
For both classical and quantum theory, this intrinsic triple-path interference is zero for any triplet of
paths. Experimental confirmation of $I_3=0$ would be a confirmation of the Born rule.  As discussed in~\cite{Huber:2021xpx} in Refs.~\cite{Sinha,Park}, bounds were placed on the parameter
\be
\kappa = \dfrac{\varepsilon}{\delta}, \quad \varepsilon =  I_3(A,B,C), \quad \delta  =  |I_2(A,B)| + |I_2(B,C)| + |I_2(C,A)|.
\ee
Ref.~\cite{Sinha} reports $\kappa=0.0064\pm 0.0119$ for a multi-slit experiment
with a single photon source, while Ref.~\cite{Park} reports $\kappa=0.007\pm 0.003$
based on a liquid state NMR experiment.
Thus, the 1$\sigma$ deviation of $\kappa$ from zero allowed by these experiments
is $|\kappa| < 0.01\sim 0.02$. (See also~\cite{Lutz1, Lutz2}.) Quite surprisingly~\cite{Huber:2021xpx}, the neutrino test of triple-interference based on the
forthcoming JUNO experiment gives the same order of precision as the electromagnetic tests.

We can easily generalize Sorkin's approach to the case of gravity.  In particular, one can consider triple-slit interference experiments (either involving photons in a three-slit experiment or 
oscillations of the three neutrino flavors) in a gravitational field. These experiments might also call for the use of gravitational or atomic interferometry as well as optical lattice atomic clocks (to which we will turn below).
The intrinsic triple correlator not present in the canonical quantum theory should be present to some degree given the above general argument on the quantum mechanical consequences of topology change in quantum gravity. In fact, the natural conjecture here is that due to an intrinsic non-linearity of
gravity one should, in principle, expect interferences of all orders in the tower of Sorkin like interferences, and also a non-linear
summation of these.

Modified quantum evolution is a typical approach in quantum gravity phenomenology --- indeed the entire area of searching for spacetime symmetry violation can be reframed as looking for quantum gravity induced modifications of propagators~\cite{Addazi:2021xuf}.  What is new  in our approach is the non-linearity which has not been significantly explored.  Beyond the toy model above, one can construct arguments directly from string theory supporting non-linear modifications.  In that case the non-linear cubic theory is string field theory (for open strings, this is the Witten cubic open string field theory~\cite{Witten:1985cc}; for closed strings there is a non-associative cubic string field theory proposed by Strominger~\cite{Strominger:1987bn} and a more canonical non-polynomial closed string field theory of Zwiebach~\cite{Zwiebach:1992ie}). String theory though has state operator correspondence, so from our new viewpoint,
the famous ``pants'' diagram of string theory (the cubic interaction vertex of string field theory) 
\cite{Polchinski:1998rq} can be viewed as an example of a non-linear modification, or gravitization, of quantum mechanics.

Furthermore this proposal can be made precise in the context of {\em\/metastring\/} theory~\cite{Freidel:2015uug, Freidel:2016pls, Freidel:2013zga, Freidel:2014qna, Freidel:2015pka, Freidel:2017xsi,  Freidel:2018apz,  Freidel:2019jor, Minic:2020oho,  Freidel:2017wst, Freidel:2017nhg, Freidel:2021wpl} (which includes a useful phenomenological limit, called the {\em\/metaparticle\/}~\cite{Freidel:2018apz}) that captures intrinsic non-commutative 
(as well as non-associative~\cite{Gunaydin:2013nqa}) features
of string theory, leading to many novel and interesting predictions~\cite{Berglund:2022qcc}. This theory can be understood as defining a precise notion of 
the {gravitization} of quantum geometry, consistent with unitarity and causality,
in the sense that quantum mechanical structures that are traditionally fixed become dynamical.
Note that the modified evolution equations discussed in the beginning of this article can be made more precise in the non-associative cubic string field theory 
\cite{Strominger:1987bn},
as well as in metastring theory~\cite{Minic:2020oho, Berglund:2020qcu}.  In fact, intrinsic triple-path interference can be directly related to the non-associativity 
\cite{Barnum:2014ysa, tripleint} which occurs in metastring theory and cubic closed string field theory, lending further credence to our phenomenological proposal.

We have so far discussed gravitizing the quantum from the quantum mechanical point of view, but there is a geometric point of view that is also insightful.  The geometry of quantum theory is very rigid and is represented by the complex projective geometry~\cite{Minic:2003en}.
Geometrically, the Born rule characterizes odd-dimensional spheres (implied by the Fisher metric in the space of distinguishable events), which in turn can be viewed as $U(1)$ bundles over complex projective spaces~\cite{Minic:2004rj}. The metric structure of the complex projective space so-specified by the Born rule
and its complex structure are compatible with the symplectic structure
 (which in a singular limit reduces to the classical symplectic structure)~\cite{Minic:2003en, Minic:2004rj}. In particular, given the Born rule this is the well known Fubini-Study metric, which is an Einstein metric and satisfies the vacuum Einstein equations with a dimension-dependent cosmological constant.  Notably, the vacuum Einstein equations are expressly not state (position) dependent --- there is no local state (position) dependent source. 

Following the work of~\cite{Minic:2003en} (and the suggestion of Ashtekar and Schilling~\cite{Ashtekar:1997ud}), 
one can ``gravitize'' this rigid geometry of quantum theory by making the Fubini-Study metric dynamical~\cite{Minic:2003en, Minic:2004rj, Jejjala:2007rn}.  
Since in the absence of state dependence the vacuum Einstein equations are satisfied, then one natural candidate for the governing dynamical equation for the Fubini-Study metric with local state dependence is simply the full quantum Einstein equation:
\be
G_{ab}[g_{cd}^{FS}]+ \Lambda\,g^{FS}_{ab}=G_{QG}\,A_{ab}(\psi)
\ee
where $A_{ab}(\psi)$ is a local function on the space of states and $G_{QG}$ is some dimensionful coupling that would vanish as we return to the Born rule.  $G_{QG}$ is technically independent of but presumably related to $G_N$, Newton's constant. While the classical and quantum theories are different (and operate in different spaces), we may consider flat-space quantum field theory with the Born rule as our unperturbed, or ``un-gravitized'' theory. Then, the vanishing of all geometrical and topological quantum gravity effects corresponds to the {\em\/double limit\/}  $G_N,G_{QG}\,{\to}\,0$, which indicates a relation between the two.   In summary, the modification to the Fubini-Study metric is in complete analogy with the transition from the
fixed Minkowski geometry of special relativity (a solution of the vacuum Einstein equations) to the dynamical geometry of general relativity where there is now a local source term.

As a final comment, we recall that the quantum clock embodies the relation between the Fubini-Study metric of the complex projective spaces and the
time interval multiplied by the uncertainty in energy (see~\cite{Minic:2003en, Minic:2004rj, Jejjala:2007rn} and the original work of Aharonov and Anandan~\cite{aa}).
The basic equation here follows from unitary evolution and reads as
$
2 \hbar\, ds_{FS} = \Delta E\, dt, 
$
where $ds_{FS}$ characterizes the Fubini-Study metric of complex projective
spaces (which is in turn directly related to the Born rule and the Fisher metric in the space of quantum states\cite{Minic:2003en}), 
and $\Delta E$ is the dispersion of energy, and $dt$ the time differential.
As we argued, in the presence of topology change the Hamiltonian becomes effectively state dependent $H \to H_{\psi}$ and so the 
energy dispersion becomes
state dependent $ \Delta E  \to (\Delta E)_{\psi}$ and thus the quantum geometry becomes dynamical $ds_{FS} \to ds_{FS} (\psi)$, and 
subject
to the above quantum Einstein equation.
What would be an experimental probe of this precise quantum statement regarding the quantum geometry?
Here we recall the recently developed optical lattice atomic clocks which have been used to measure the gravitational red shift to a remarkable accuracy 
\cite{katori}. In this case one tests corrections to the red-shift formula $\frac{\Delta \nu}{\nu} = (1 + \alpha) \frac{\Delta U}{c^2}$,
where $\nu$ is the frequency and $U$ the gravitational potential, and the correction factor $\alpha$ (equal to zero in general relativity)
is bounded to the order of $10^{-5}$.
In view of the relation between quantum geometry and quantum clocks, the ultra-precise experimental tests of the validity of general relativity can be used 
in principle to constrain the gravitization of quantum geometry.

In summary: classical (Einsteinian) gravity gravitizes all of classical physics by making all geometric structures dynamical. In this essay we have discussed a new approach 
to the problem of quantum gravity, 
based on the idea of gravitizing the quantum, according to which
traditionally fixed quantum mechanical structures  become dynamical. We discussed the general reason behind this
expectation and then we addressed specific ways of testing this 
new approach to quantum gravity via multi-path interferences and the experimental probes of the geometry
of quantum theory using optical lattice atomic clocks. This new approach should also have important theoretical and experimental 
consequences for more traditional areas of quantum gravity
such as quantum cosmology and quantum black holes.

\paragraph{Acknowledgments:} We thank Laurent Freidel, Jerzy Kowalski-Glikman, Rob Leigh and Tatsu Takeuchi for
inspiring discussions. 
PB would like to thank the CERN Theory Group
for their hospitality over the past several years. TH is grateful to the Department of Physics,
University of Maryland, College Park MD, and the Physics Department of the Faculty of
Natural Sciences of the University of Novi Sad, Serbia, for the recurring hospitality and resources.
D. Minic is grateful to Perimeter Institute for hospitality and support. The work of PB and D. Mattingly
is supported in part by the Department of Energy grant DE-SC0020220.
D. Minic thanks the Julian Schwinger Foundation and the U.S. Department
of Energy under contract DE-SC0020262 for support. 
This work contributes to the European Union COST Action CA18108 Quantum gravity phenomenology in the multi-messenger approach.

\end{document}